\documentclass{article}

\usepackage{amssymb,amsfonts,amsmath}
\usepackage{cite,enumerate,float}
\usepackage{color}
\usepackage{tikz}
\usetikzlibrary{arrows,snakes,backgrounds}
\usepackage{mathtools,leftindex }
\usepackage{xcolor}
\definecolor{airforceblue}{rgb}{0.36, 0.54, 0.66}	\definecolor{cobalt}{rgb}{0.0, 0.28, 0.67}
\definecolor{cyan(process)}{rgb}{0.0, 0.72, 0.92}
\usepackage[pdftex,unicode,
colorlinks=true,
linkcolor = blue,
citecolor = cyan(process),
urlcolor = airforceblue]{hyperref}

\def\be{\begin{eqnarray}}
\def\ee{\end{eqnarray}}
\def\nn{\nonumber}

\def\e{{\bf e}}
\def\f{{\bf f}}
\def\E{{\bf E}}
\def\F{{\bf F}}
\def\h{{\bf h}}
\def\x{{\bf x}}

\definecolor{red}{rgb}{1,0,0}
\definecolor{orange}{rgb}{1,0.5,0}
\definecolor{violet}{rgb}{0.7,0,1}

\def\C{{\cal C}}

\hoffset= - 1.0in         % switch off for draft style
\begin{document}

\title{\vspace{1.5cm}\bf
Tau-functions beyond the group elements
}

\author{
A. Mironov$^{b,c,d,}$\footnote{mironov@lpi.ru,mironov@itep.ru},
V. Mishnyakov$^{a,b,c,e,}$\footnote{mishnyakovvv@gmail.com},
A. Morozov$^{a,c,d,}$\footnote{morozov@itep.ru}
}

\date{ }

\maketitle

\vspace{-6.5cm}

\begin{center}
\hfill FIAN/TD-15/23\\
\hfill IITP/TH-21/23\\
\hfill ITEP/TH-27/23\\
\hfill MIPT/TH-20/23
\end{center}

\vspace{4.5cm}

\begin{center}
$^a$ {\small {\it MIPT, Dolgoprudny, 141701, Russia}}\\
$^b$ {\small {\it Lebedev Physics Institute, Moscow 119991, Russia}}\\
$^c$ {\small {\it NRC ``Kurchatov Institute", 123182, Moscow, Russia}}\\
$^d$ {\small {\it Institute for Information Transmission Problems, Moscow 127994, Russia}}\\
$^e$ {\small{\it Institute for Theoretical and Mathematical Physics, Lomonosov Moscow State University, Moscow 119991, Russia}}
\end{center}

\vspace{.1cm}

\begin{abstract}
Matrix elements in different representations are connected by quadratic relations.
If matrix elements are those of a {\it group element}, i.e. satisfying the property
$\Delta(X) = X\otimes X$, then their generating functions
obey bilinear Hirota equations and hence are named $\tau$-functions.
However, dealing with group elements is not always easy,
especially for non-commutative algebras of functions, and this slows down the development of $\tau$-function theory
and the study of integrability properties of non-perturbative functional integrals.
A simple way out is to use arbitrary elements of the universal enveloping algebra, and not just the group elements.
Then the Hirota equations appear to interrelate a whole system of generating functions,
which one may call {\it generalized} $\tau$-functions.
It was recently demonstrated that this idea  can be applicable even  to a somewhat
sophisticated case of the quantum toroidal algebra.   We consider a number of simpler examples, including ordinary and quantum groups,  to explain how the method works and what kind of solutions one can obtain.
\end{abstract}

\bigskip

\section{Introduction}

$\tau$-functions play the prominent role in modern theoretical physics.
It turns out that non-perturbative partition functions of quantum theories,
as functions of coupling constants and boundary conditions,
belong to this class and satisfy the peculiar type of Hirota bilinear relations
\cite{Hirota,DJKM1,DJKM,Kac,UT,GKLMM,KMM,MMV,gentau,UFN23}.
To make this statement general enough one needs a broad definition of $\tau$-functions,
not restricted to conventional Toda-KP family.
That one is actually associated with the fundamental representations of $\mathfrak{sl}_\infty$
algebra, but bilinear relations are far more general and remain true for
representation theory of arbitrary universal enveloping algebras (UEA) ${\cal G}$ \cite{GKLMM,KMM,gentau}.

More concretely, one considers a group element, i.e. an element $g\in {\cal G}\otimes{\cal A}$, where ${\cal A}={\cal G}^*$ is the algebra of functions, such that its comultiplication $\Delta(g)=g\otimes g$, and realizes that there is a set of bilinear relations for its matrix elements, $\langle\vec m|g|\vec n\rangle$.
Defining the $\tau$-function as a generating function of all these matrix elements,
\be
\tau(t,\bar t)=\sum_{\vec m,\vec n}\langle\vec m|g|\vec n\rangle t_{\vec m}\bar t_{\vec n}
\ee
where $t_{\vec m}$ and $\bar t_{\vec n}$ are sets of the generating parameters,
one rearranges this set of bilinear relations into differential or difference equations for this $\tau$-function w.r.t. $t$ and $\bar t$.

Note that there are, however, two problems \cite{GKLMM,KMM,gentau}: first of all, such $\tau$-functions, while satisfying quadratic Hirota-like equations are not necessarily related to integrability in its usual sense of commuting flows.
Commutativity requires a special care:
in generic representations, the flows in $t$ and $\bar t$ generating all matrix elements are not commutative (which kills integrability).

There is however, a more severe problem, that with the very existence of bilinear equations.
If one deals with the quantum deformation of UEA, matrix elements of the group elements (i.e. elements of the algebra of functions ${\cal A}$) are not commutative, since the comultiplication is not co-commutative.
At the same time, usage of the group element is necessary, since it makes the equations closed.

A somewhat radical idea, which recently got a new momentum in \cite{Bourg},
is to lift restricting to the group element, and consider
{\it generalized} $\tau$-functions associated with arbitrary elements
of $X\in {\cal G}$ (i.e. $X$ is no longer an element of ${\cal G}\otimes{\cal A}$),
generally not satisfying the property $\Delta(X)=X\otimes X$.
Then the bilinear equations do not close on a particular generating function
but interrelate many of them.
However, one can easily make the emerging structure {\it hierarchical}
and {\it convergent}, thus making the new extended system of $\tau$-functions
and Hirota bilinear equations conceptually as simple as the conventional one. This allows one to solve, at the least, the second problem: one can now successfully work in the cases of quantum deformations, when the comultiplication is not co-commutative. In particular, one can deal with special representations of even $q,t$-deformed UEA like the Ding-Iohara-Miki (DIM) or quantum toroidal algebra \cite{DIM}.

In this paper, we go through a number of examples, and show how this approach
works in various systems. In section 2, we describe two equivalent but technically different approaches to constructing Hirota bilinear equations. In sections 3 and 4, we consider a series of examples the bilinear equations for generalized $\tau$-functions (i.e. those associated not obligatory with the group element), starting from the simplest case of ${\cal G}=U(\mathfrak{sl}_2)$, and then describe ${\cal G}=U_q(\mathfrak{sl}_2)$, ${\cal G}=U(\mathfrak{sl}_N)$ and, finally, very sketchy, the quantum toroidal algebra, the example described in detail in \cite{Bourg}. In section 5, we discuss the notion of universal $\tau$-functions introduced in \cite{KMM,gentau} and, a bit differently, in \cite{Bourg}, while some concluding remarks can be found in section 6.

\section{Two approaches to bilinear equations}

There are two ways to derive bilinear equations for the $\tau$-function. One is due to \cite{DJKM} and implies a construction using operators intertwining various representations of ${\cal G}$ \cite{GKLMM,KMM,gentau}. It gives rise to bilinear equations relating $\tau$-functions in various, in particular, distinct representations, and, in the case of ${\cal G}=U(\mathfrak{sl}_\infty)$ and of fundamental representations, it is the celebrated Toda chain equation
\be
\partial\tau_n\bar\partial\tau_n-\tau_n\partial\bar\partial\tau_n=\tau_{n+1}\tau_{n-1}
\ee
where the subscript $n$ refers to the $n$-th fundamental representation of $\mathfrak{sl}_\infty$, and $\partial:={\partial\over\partial t_1}$, $\bar\partial:={\partial\over\partial \bar t_1}$ are the derivatives w.r.t. generating parameters associated with the sum of all simple roots generators, positive and negative correspondingly.

The second way to derive bilinear identities is due to \cite{KW}, and it uses the fact that the split Casimir operator, i.e. the Casimir operator in the square of representation commutes with the comultiplication \cite{splitC,Isaev}. This way, one can get bilinear equations for the $\tau$-function in this representation.

\subsection{Intertwining operator trick\label{inttw}}

The derivation of the bilinear identities that generalizes the standard derivation using the fermionic language \cite{DJKM}.

The starting point of the derivation \cite{GKLMM,KMM,gentau} is to embed a Verma module
$\widehat V$ into the tensor product $V\otimes W$, where
$W$ is some (arbitrary) finite-dimensional representation of
${\cal G}$. With the fixed choice of $V$ and $W$, there exist only
finite number of possible $\widehat V$.

Now one defines intertwining operators, which generalize the notion of fermions.
The right intertwining operator of the type $W$ is defined to be a homomorphism of the
${\cal G}$-modules:
\be\label{inttw}
E_R: \ \ \widehat V \longrightarrow V\otimes W
\ee

Similarly, one considers another triple of modules that define
the left intertwining operator
\be\label{inttw2}
E'_L: \ \ \widehat V' \longrightarrow W' \otimes V'
\ee
so that the product $W\otimes W'$ contains the {\it
unit} representation of ${\cal G}$. These two constructed intertwining operators generalize the notion of fermions.

Now one considers the projection
of the product $W\otimes W'$ onto the unit representation
\be\label{proj}
\pi: \ \ W\otimes W' \longrightarrow I
\ee
Using this projection, one can construct the new intertwining operator
\be\label{Gamma}
\Gamma: \ \
\widehat V \otimes \widehat V' \stackrel{E_R\otimes E_L'}{\longrightarrow}
V \otimes W \otimes W' \otimes V'
\stackrel{I\otimes \pi \otimes I}{\longrightarrow} V \otimes V'
\ee
possessing the property
\be
\Gamma (g\otimes g) = (g\otimes g) \Gamma
\label{Ggg=ggG}
\ee
for any group element $g$ such that
\be\label{grel}
\Delta(g)=g\otimes g
\ee

Put it differently, the space $W\otimes W'$ contains the canonical
element of pairing $w_i\otimes w^i$ commuting with the action of
$\Delta (g)$. This means that the operator
$\sum_i E_i\otimes E^i :  V\otimes V' \longrightarrow \widehat V\otimes
\widehat V'$ ($E_i\equiv E(w_i),\ E^i\equiv E(w^i)$) commutes with $\Delta
(g)$.

Now taking the matrix elements of identity (\ref{Ggg=ggG}), one obtains
\be
\Big\langle\Psi_1\Big|\Gamma (g\otimes g)\Big|\Psi_2\Big\rangle =
\Big\langle\Psi_1\Big| (g\otimes g) \Gamma\Big| \Psi_2\Big\rangle
\ee
where $\Big\langle\Psi_{1}\Big|$ and $\Big|\Psi_{2}\Big\rangle$ are arbitrary vectors, and we choose them to be
\begin{equation}
	\begin{split}
		&\Big\langle\Psi_1\Big|=\Big\langle 0\Big|E_t\otimes \Big\langle 0\Big|E_{t'}\\
		&\Big|\Psi_2\Big\rangle=E_{\bar t}\Big|0\Big\rangle_\lambda\otimes E_{\bar t'}\Big|0\Big\rangle_{\lambda'}
	\end{split}
\end{equation}
where $\Big|0\Big\rangle_\lambda$ denotes the highest weight vector in the representation $\lambda$, and $E_t$, some properly chosen function of generators, see examples below. In order to obtain  the Hirota bilinear equations, one has to use the commutation relations of the intertwining operators with algebra generators and push $E_i$ to the highest weight vector. The
result of this procedure can be imitated by the action of some differential
or difference operators that leads to the differential or difference Hirota bilinear identities for the $\tau$-function
\be
\tau_\lambda(t,\bar t;g)=\prescript{}{\lambda}{\Big\langle} 0\Big|E_t\cdot g\cdot E_{\bar t}\Big|0\Big\rangle_\lambda
\ee
This latter step is, however, not always possible and requires the correct choice
of the function $E_t$.

In particular, bilinear equations of the Toda/KP hierarchy are obtained with the choice of ${\cal G}=U(\mathfrak{sl}_\infty)$, and the triples: $\widehat V=F_{k+1}$, $V=F_k$, $W=F_1$, and $\widehat V'=F_{k-1}$, $V'=V=F_k$, $W'=F_1^*$, where $F_k$ denotes the $k$-th fundamental representation, and $F_k^*$, its conjugate. This gives rise \cite{KMM} to the standard Hirota bilinear equations of the Toda lattice hierarchy for the $\tau$-function $\tau_k(t,\bar t;g)$ \cite{UT,DJKM}.

\subsection{The ``split Casimir'' trick}

 This long story from \cite{GKLMM} can be made shorter and sometimes easier generalizable by the trick \cite{KW} which is nowadays often called the split Casimir approach \cite{splitC,Isaev} (it is just a nickname for the comultiplication action on the Casimir operator)\footnote{A similar way to derive differential/difference equation by Casimir operator insertions was also proposed in \cite{GKMMMO}.}. This way to derive the same bilinear Hirota equations is obtained from the evident relation
\be\label{sCe}
\Big\langle\Psi_1\Big|\Delta(\C_2)\Delta(X)\Big|\Psi_2\Big\rangle =
\Big\langle\Psi_1\Big| \Delta(X)\Delta(\C_2)\Big| \Psi_2\Big\rangle
\ee
where $\C_2$ is the second Casimir operator\footnote{One can use higher Casimir operators, however, it would just lead to higher order differential equations.}, and $X$ is an arbitrary element of the UEA. Now one can push $\Delta(\C_2)$ to the vector $\prescript{}{\lambda}{\Big\langle} 0\Big|$ at the l.h.s. of (\ref{sCe}), and to the vector $\Big|0\Big\rangle_{\lambda}$ at the r.h.s. of (\ref{sCe}). Again, the
result of this procedure can be imitated by the action of some differential
or difference operators that leads to the differential or difference bilinear identities for the $\tau$-function
\be
\tau_\lambda(t,\bar t;X)=\prescript{}{\lambda}{\Big\langle} 0\Big|E_t\cdot X\cdot E_{\bar t}\Big|0\Big\rangle_\lambda
\ee

In this note, we are going to consider mainly a generic case when $\Delta(X) = \sum X'\otimes X'' \neq X\otimes X$, then
they include the whole set of $\tau$-functions
$\tau(\bar t, t|X) := \Big\langle E_{\bar t} X E_t\Big\rangle$ with different $X=X',X''$.
However, in this case:

\begin{itemize}
\item{}This allows us not to restrict the coefficients of $X$ to belong to the
algebra of functions, which is non-commutative in the case of quantum deformations.

\item{} Moreover, the set of solutions is at least as big as before the quantum deformations:
each $X$ allowed for the non-deformed $\tau$ continues to be allowed.

\item{} In fact, this set is much  bigger: now there is a solution for arbitrary $X$
from the universal enveloping algebra, not just a linear exponential of generators.
\end{itemize}

In the next section, as an illustration, we consider bilinear equations for the $\tau$-functions associated with arbitrary elements of UEA and obtained within the split Casimir approach, while, in section 4, we discuss examples of bilinear equations for such $\tau$-functions obtained using the intertwining operators.

\section{Basic examples: $U(\mathfrak{sl}_2)$ and $U_q(\mathfrak{sl}_2)$}

\subsection{$U(\mathfrak{sl}_2)$ algebra\label{sl2}}

To illustrate these ideas, we start with the simplest example of ${\cal G}=U(\mathfrak{sl}_2)$, and deal with its arbitrary highest weight representation with the highest weight $\lambda$: in this case, there is no reason to restrict oneself to the fundamental representations only.

The commutation relations of the $\mathfrak{sl}_2$ Lie algebra are
\be
\phantom{.}[\e ,\f ]=\h\nn\\
\phantom{.}[\h ,\e ]=2\e\\
\phantom{.}[\h ,\f ]=-2\f\nn
\ee
In this case, the $\tau$-function, which is the generating function of matrix elements at any representation $\lambda$ is defined
\be\label{taucl}
\tau_\lambda(t;X) = {}_\lambda \langle 0|e^{t\e}Xe^{\bar t\f}|0\rangle_\lambda
\ee
where $X$ an element of $U(\mathfrak{sl}_2)$, $|0\rangle_\lambda$ is the highest weight vector of representation $\lambda$,
\be\label{irrep}
\e |0\rangle_\lambda=0,\ \ \ \ \ \ {}_\lambda \langle  0|\f =0,\ \ \ \ \ \ \h |0\rangle_\lambda=\lambda |0\rangle_\lambda
\ee
The comultiplication is
\be
\Delta(\x)=\x\otimes I+I\otimes \x
\ee
where $\x$ is any element of $\e$, $\f$, $\h$.

When $X$ is a group element, i.e. an element of the UEA with the comultiplication $\Delta(X_g)=X_g\otimes X_g$,
the bilinear equations can be derived in the first way (from intertwining operators), in this case, they are \cite[Eqs.(34)-(35)]{GKLMM}
\be\label{eq0}
\Big(\lambda\bar\partial'-\lambda'\bar\partial+(\bar t-\bar t')\bar\partial\bar\partial'\Big)\tau_{\lambda}(t;X_g)
\tau_{\lambda'}(t';X_g)=\lambda\lambda'(t-t')\tau_{\lambda-1}(t;X_g)
\tau_{\lambda'-1}(t';X_g)
\ee
and
\be
\Big(\lambda'+(\bar t-\bar t')\bar\partial'\Big)\tau_{\lambda}(t;X_g)
\tau_{\lambda'}(t';X_g)={\lambda'\over\lambda+1}\Big(\lambda+1-(t-t')\partial\Big)
\tau_{\lambda+1}(t;X_g)
\tau_{\lambda'-1}(t';X_g)
\ee
where $\partial={\partial\over\partial t}$, $\bar \partial={\partial\over\partial \bar t}$. There are definitely many more bilinear equations (e.g. \cite[Eq.(36)]{GKLMM}).

When $X$ is not restricted to be a group element, one has to use the Kac-Wakimoto approach. In order to use it, one notes that the simplest split Casimir operator $\C$ in this case is ($\C_2$ is the second Casimir operator)
\be\label{sC}
\C=\Delta(\C_2)-I\otimes \C_2-\C_2\otimes I=\e\otimes \f+\f\otimes \e+{1\over 2}\h\otimes \h
\ee
and it commutes with the comultiplication:
\be
\Delta(X)\C=\C\Delta(X),\ \ \ \ \ \forall X\in U_{\mathfrak{sl}_2}
\ee
In (\ref{sC}), we subtracted from $\Delta(\C_2)$ a non-split part, which trivially commutes with the comultiplication.

Now one can take the average of this identity of the form:
\be
0={}_{\lambda} \langle  0|e^{t\e}\otimes{}_{\lambda'} \langle  0|e^{t'\e}\Delta(X)\C e^{\bar t\f}|0\rangle_\lambda\otimes e^{\bar t'\f}|0\rangle_{\lambda'}-
{}_{\lambda} \langle  0|e^{t\e}\otimes{}_{\lambda'} \langle  0|e^{t_1'\e}\C\Delta(X)e^{\bar t\f}|0\rangle_\lambda\otimes e^{\bar t'\f}|0\rangle_{\lambda'}
\ee
and pull $\C$ in the first term up to the right highest vectors, and in the second term, to the left\footnote{
We used that
\be
\e\exp(t\f)=\exp(t\f)(\e+t\h-t^2\f)\nn\\
\exp(t\e)\f=(\f+t\h-t^2\e)\exp(t\e)\nn\\
\h\exp(t\f)=\exp(t\f)(\h-2t\f)\nn\\
\exp(t\e)\h=(\h-2t\e)\exp(t\e)\nn
\ee
}:
\be\label{i1}
0={}_{\lambda} \langle  0|e^{t\e}\otimes{}_{\lambda'} \langle  0|e^{t'\e}\Delta(X)|V_1\rangle-\langle V_2|\Delta(X)e^{\bar t\f}|0\rangle_\lambda\otimes e^{\bar t'\f}|0\rangle_{\lambda'}
%\nn\\
\ee
{\footnotesize
\be
|V_1\rangle:=(\lambda \bar t-\bar t^2\f)e^{\bar t\f}|0\rangle_\lambda\otimes\f
e^{\bar t'\f}|0\rangle_{\lambda'}+\f e^{\bar t\f}|0\rangle_\lambda\otimes (\lambda'\bar t'-\bar t'^2\f)e^{\bar t'\f}|0\rangle_{\lambda'}
%+\nn\\
+{1\over 2}(\lambda-2\bar t\f)e^{\bar t\f}|0\rangle_\lambda\otimes(\lambda'-2\bar t'\f )e^{\bar t'\f}|0\rangle_{\lambda'}\nn\\
\langle V_2|:={}_{\lambda} \langle  0|e^{t\e}\e\otimes{}_{\lambda'} \langle  0|e^{t'\e}(\lambda't'-t'^2\e)+{}_{\lambda} \langle  0|e^{t\e}(\lambda t-t^2\e)\otimes{}_{\lambda'} \langle  0|e^{t'\e}\e+
{1\over 2}{}_{\lambda} \langle  0|e^{t\e}(\lambda-2t\e)\otimes{}_{\lambda'} \langle  0|e^{t'\e}(\lambda'-2t\e)\nn\\
\nn
\ee
}
Now we do not specify $X$ to be a group element, just put
\be\label{cop}
\Delta(X)=\sum_\alpha X_\alpha'\otimes X_\alpha''
\ee
and reproduce the action of generators in (\ref{i1}) by differential operators.
Then, one finally obtains from (\ref{i1}) the bilinear equation
\be\label{besl2}
\Big[(t-t')^2\partial\partial'-(t-t')(\lambda\partial'-\lambda'\partial)-(\bar t-\bar t')^2\bar\partial\bar\partial'+(\bar t-\bar t')(\lambda\bar\partial'-\lambda'\bar\partial)\Big]
\sum_\alpha\tau_\lambda(t;X_\alpha')\tau_{\lambda'}(t';X_\alpha'')=0
\ee
Let us now choose $\lambda'=\lambda$ and expand this equation in $t-t'$ in order to get the pair of bilinear relations
\be\label{eq1}
\sum_\alpha\Big[(\lambda-1)\partial\tau_\lambda(t;X_\alpha')\partial\tau_\lambda(t;X_\alpha'')-
\lambda\tau_\lambda(t;X_\alpha')\partial^2\tau_\lambda(t;X_\alpha'')\Big]=0
\ee
and similarly for the expansion in $\bar t-\bar t'$:
\be\label{eq2}
\sum_\alpha\Big[(\lambda-1)\bar\partial\tau_\lambda(t;X_\alpha')\bar\partial\tau_\lambda(t;X_\alpha'')-
\lambda\tau_\lambda(t;X_\alpha')\bar\partial^2\tau_\lambda(t;X_\alpha'')\Big]=0
\ee

Let us consider an example of $X=X_g$ and of $X_e$ such that the sum in (\ref{cop}) consists of two terms, the typical element being $X=\x$. Denote $\tau_g^{(\lambda)}:=\tau_\lambda(t;X_g)$, $\tau_x^{(\lambda)}:=\tau_\lambda(t;X_e)$. Then, equation (\ref{eq1}) becomes
\be\label{eq11}
(\lambda-1)\Big(\partial\tau_g^{(\lambda)}\Big)^2-
\lambda\tau_g^{(\lambda)}\partial^2\tau_g^{(\lambda)}=0\nn\\
2(\lambda-1)\partial\tau_g^{(\lambda)}\partial\tau_x^{(\lambda)}-
\lambda\Big(\tau_x^{(\lambda)}\partial^2\tau_g^{(\lambda)}+\tau_g^{(\lambda)}\partial^2\tau_x^{(\lambda)}\Big)=0
\ee
and (\ref{eq2}) becomes
\be\label{eq22}
(\lambda-1)\Big(\bar\partial\tau_g^{(\lambda)}\Big)^2-
\lambda\tau_g^{(\lambda)}\bar\partial^2\tau_g^{(\lambda)}=0\nn\\
2(\lambda-1)\bar\partial\tau_g^{(\lambda)}\bar\partial\tau_x^{(\lambda)}-
\lambda\Big(\tau_x^{(\lambda)}\bar\partial^2\tau_g^{(\lambda)}+\tau_g^{(\lambda)}\bar\partial^2\tau_x^{(\lambda)}\Big)=0
\ee
The first equations in (\ref{eq11}), (\ref{eq22}) have a solution
\be\label{sol1}
\tau_g^{(\lambda)}=\alpha_g\Big(1+C_1t+C_2\bar t+C_3t\bar t\Big)^\lambda
\ee
There are 4 arbitrary parameters since this equation is correct for any group element $X_g$, and the group elements are parameterized by 3 parameters, and there is also a normalization factor. Note that Eq.(\ref{eq0}) has the same solution (\ref{sol1}). One can easily check that this solution satisfies the full set of equations (\ref{besl2}) so that  (\ref{eq11}), (\ref{eq22}) turns out to be enough for evaluating $\tau_g^{(\lambda)}$.

Similarly, the second equations in (\ref{eq1}), (\ref{eq2}) have a solution
\be\label{sol2}
\tau_x^{(\lambda)}=\alpha_x\lambda\Big(1+A_1t+A_2\bar t+A_3t\bar t\Big)\Big(1+C_1t+C_2\bar t+C_3t\bar t\Big)^{\lambda-1}
\ee
Here the 3 parameters $A_i$ parameterize elements $X$ of the UEA such that the comultiplication $\Delta(X)=X_g\otimes X_1+X_1\otimes X_g$. These are elements of the form\footnote{One can also write $\Delta(g)\sim g\otimes g+g\otimes g$, which gives rise to 1 in the first multiplier in (\ref{sol2}).} $X_e=X_g(\e\oplus \f\oplus \h)$, and they can be obtained from the $\tau$-function (\ref{sol1}) by derivatives in parameters $C$'s, which parameterize the group element. In this sense, (\ref{sol2}) and (\ref{sol1}) are not quite independent.

One can now check that the complete equation (\ref{besl2}) is solved by the pair of $\tau$-functions (\ref{sol1}) and (\ref{sol2}) (to this end, it is necessary to separate the factor $\lambda$ in (\ref{sol2}) from the arbitrary constant $\alpha_x$).

After considering $X_g$ and $X_x$, one can further consider $X_{xx}$, the element with a typical representative $\x^2$, etc. Each new element gives rise to a new equation, but it seems to exist a clear hierarchy: there is a closed equation for $\tau^{(\lambda)}_g$, there are two equations for $\tau^{(\lambda)}_x$ involving also that for $\tau^{(\lambda)}_g$, etc. All these $\tau$-functions can be generated by multiple derivatives of $\tau_g^{(\lambda)}$ in parameters $C$'s that parameterize the group element (algebra of functions), and, hence, in a sense, {\bf they do not give rise to new independent hierarchies!} One can say they are associated with infinitesimal invariant B\"acklund transforms.

\subsection{$U_q(\mathfrak{sl}_2)$ algebra}

Consider the $q$-deformation of the UEA of $\mathfrak{sl}_2$: the $U_q(\mathfrak{sl}_2)$ algebra is given by the defining relations
\be
\phantom{.}[\e,\f]={q^\h-q^{-\h}\over q-q^{-1}}\nn\\
\phantom{.}q^\h\e=q^2\e q^\h\nn\\
\phantom{.}q^\h\f=q^{-2}\f q^\h
\ee
we choose the comultiplication
\be\label{qcomul}
\Delta(\e)=q^{\h/2}\otimes\e+\e\otimes q^{-\h/2}\nn\\
\Delta(\f)=q^{\h/2}\otimes\f+\f\otimes q^{-\h/2}\nn\\
\Delta(q^\h)=q^\h\otimes q^\h
\ee
and the second Casimir operator is
\be
\C_2=\e\f+{q^{-1}q^\h+qq^{-\h}\over (q-q^{-1})^2}
\ee
i.e. the split Casimir operator is
\be
\C=\Delta(\C_2)=q^\h\otimes\e\f+\e\f\otimes q^{-\h}+\e q^{\h/2}\otimes q^{-\h/2}\f+q^{\h/2}\f\otimes\e q^{-\h/2}+
{q^{-1}q^\h\otimes q^\h+qq^{-\h}\otimes q^{-\h}\over (q-q^{-1})^2}
\ee
and there is no trivial part at the r.h.s.: all terms are split.

The highest weight representation is the same as in the non-deformed case, hence, formula (\ref{irrep}) preserves. We now choose the following definition of the $\tau$-function
\be\label{tauq}
\tau_\lambda(t;X;q)={}_{\lambda} \langle  0|e_{q}(t\e)Xe_{q^{-1}}(\bar t\f)|0\rangle_\lambda
\ee
where $e_q(x)$ denotes the $q$-exponential\footnote{Note that this definition corresponds to the replace $q\to q^2$ as compared with the standard definition.} \cite{qexp},
\be
e_q(x):=\sum_{n\ge 0}{x^n\over [n]!}q^{-n(n-1)/2}
\ee
$[n]=(q^x-q^{-x})/(q-q^{-1})$ is the $q$-number, and $X\in U_q(\mathfrak{sl}_2)$. The $q$-exponential enjoys the property
$D_{x,q}e_q(ax)=ae_q(ax)$, where
\be
D_{x,q}f(x):={f(q^2x)-f(x)\over (q^2-1)x}
\ee

Performing the calculations of the previous subsection in this case (see also \cite{MV}), one obtains instead of (\ref{besl2}), the bilinear identities\footnote{We used that (the map that interchanges $\f$ and $\e$ is associated with the transform $q\to q^{-1}$, $h\to -h$)
\be
\phantom{.}[\e,e_{q^{-1}}(t\f)]=t{e_{q^{-1}}(t\f)q^\h-q^{-\h}e_{q^{-1}}(t\f)\over q-q^{-1}}\nn\\
\phantom{.}[e_{q}(t\e),\f]=t{e_{q}(t\e)q^\h-q^{-\h}e_{q}(t\e)\over q-q^{-1}}\nn\\
q^{\pm\h}e_{q^{-1}}(t\f)=e_{q^{-1}}(q^{\mp 2}t\f)q^{\pm\h}\nn\\
e_{q}(t\e)q^{\pm\h}=q^{\pm\h}e_{q}(q^{\mp 2}t\e)\nn
\ee
}
\be\label{sl2qeq}
\left[\Big({q\over q-q^{-1}}M^{-2}M'^{-2}+t'M^{-2}M'^{-2}D'+tM^{-2}M'^{2}D-q^{-\lambda'}t'M^{-2}D'
-q^{-\lambda}tM'^{2}D+\right.\nn\\
+{q^{-1}\over q-q^{-1}}M^{2}M'^{2}
+t'M^{-1}M'^{-1}D-q^{-\lambda'}t'M^{-1}M'D-q^{-\lambda}tM^{-1}M'D'+tM^{-3}M'D'
\Big)-
\nn\\
-\Big(q^{\lambda'}\bar t'\bar M^{-2}\bar D'+{q\over q-q^{-1}}\bar M^{-2}\bar M'^{-2}+
{q^{-1}\over q-q^{-1}}\bar M^2\bar M'^2-\bar t\bar M^2\bar M'^2\bar D+q^\lambda \bar t\bar M'^2\bar D
-\nn\\
\left.
-\bar t'\bar M^{-2}\bar M'^2\bar D'-\bar t'\bar M^{-1}\bar M'^{3}\bar D-\bar t\bar M\bar M'\bar D'
+q^{\lambda'}\bar t'\bar M^{-1}\bar M'\bar D+q^\lambda \bar t\bar M^{-1}\bar M'\bar D'
\Big)\right]\times\nn\\
\times\sum_\alpha\tau_\lambda(t;X_\alpha';q)\tau_{\lambda'}(t';X_\alpha'';q)=0
\ee
where
\be
M_xf(x)=f(qx)
\ee
and we denoted $D:=D_{t,q}$, $\bar D:=D_{\bar t,q^{-1}}$, $M:=M_tq^{-\lambda/2}$, $\bar M:=M_{\bar t}q^{-\lambda/2}$, and similarly for $M'$, $D'$.

\bigskip

Solutions of this difference equation can be classified by the number of terms in the sum over $\alpha$,
and it is especially instructive to see how the standard answer at $q=1$ is recovered.
For this purpose, it is convenient to introduce a new notion: the {\it rank} of the solution,
equal to the number of terms in the sum over $\alpha$.
For every {\it rank}, there will be solutions which have the right limit at $q=1$, but they will split in
several ``families".

\bigskip

Consider first the case when the sum over $\alpha$ consists of just {\bf one} term, like we did for $q=1$.

Naively,  $\tau_\lambda(t;X_g;q)\tau_{\lambda'}(t';X_g;q)$ is a solution only if
\be\label{sl2q}
\tau_\lambda(t;X_g;q)=\tau_{g,1}^{(\lambda)}(q)=\alpha_g\prod_{i=1}^{\lambda}\Big(1+Ct\bar t q^{\lambda-2i+1}\Big)=
%\alpha_g\sum_{i=0}^\lambda{[\lambda]!\over[\lambda-i]![i]!}\Big(Ct\bar t\Big)^i=
\alpha_g\sum_{i=0}^\lambda\binom{\lambda}{i}_q\Big(Ct\bar t\Big)^i:=\alpha_g\Big[(1+Ct\bar t)^\lambda\Big]_q
\ee
for integer\footnote{The last sum can be extended to non-integer $\lambda$ as well \cite{GKLMM}:
\be
\tau_\lambda(t;X_g;q)=\alpha_g\sum_{i=0}{\Gamma_q(\lambda+1)\over\Gamma_q(\lambda-i+1)\Gamma_q(i+1)}\Big(Ct\bar t\Big)^i
\nn
\ee
where $\Gamma_q(x)$ is the $q$-$\Gamma$-function \cite{qexp}. Notice that we use the slightly different definition of $q$-number $[n]=(q^n-q^{-n})/(q-q^{-1})$ instead of $(q^n-1)/(q-1)$, and the $q$-factorial is accordingly defined a bit differently.} $\lambda$, where the $q$ binomial coefficients are $\binom{\lambda}{i}_q={[\lambda]!\over[\lambda-i]![i]!}$, and the $q$-factorial $[n]!=\prod_{i=1}^n[i]$. This expression is a $q$-deformation of the Newton binomial expansion of $(1+Ct\bar t)^\lambda$, we used for it the notation $[1+Ct\bar t]^\lambda$ in \cite{GKLMM}.
What  we observe is that there are less solutions than in the non-deformed case: it is only the two-parametric solution
(parameters $\alpha_g$ and $C$) instead of the 4-parametric one in (\ref{sol1}).
This is because now the group elements are only those associated with the Cartan element.

However, in this single-$\alpha$ ({\it rank}-one) case, there is also another family of solutions.
Namely, the product $\tau_\lambda(q^{-\lambda/2}t;X_0;q)\tau_{\lambda'}(q^{\lambda'/2}t';X_0;q)$ solves the equation (\ref{sl2qeq}) with
\be\label{tauX0}
\tau_\lambda(t;X_0;q)=\tau_{g,2}^{(\lambda)}(q)=
\alpha_g \left\{\sum_{i=0}^\lambda q^{-i(i-1)/2}\binom{\lambda}{i}_q(C_1t)^i\right\}\times
\left\{\sum_{j=0}^\lambda q^{j(j-1)/2}\binom{\lambda}{j}_q(C_2\bar t)^j\right\}
\ee
The element $X_0$ is discussed in the next subsection.

Together (\ref{sl2q}) and (\ref{sl2qeq}) have the same number of parameters that a single family had at $q=1$
(2+3-1=4,
the normalization $\alpha_g$ should not be counted twice).

\bigskip

The situation is quite the same in the case of {\it rank} $2$.
One can solve the equation for $X_x$ associated with {\bf two} terms in the comultiplication sum, e.g. that for elements $X_g\cdot\e$ and $X_g\cdot\f$, i.e. for $\tau_\lambda(t;X_gq^{\h/2};q)\tau_{\lambda'}(t';X_x;q)+\tau_\lambda(t;X_x;q)\tau_{\lambda'}(t';X_gq^{-\h/2};q)$ in (\ref{sl2qeq}) with $\tau_\lambda(t;X_g;q)$ given by (\ref{sl2q}) (because of the comultiplication rule (\ref{qcomul})). Since $\tau_\lambda(t;X_gq^{\pm\h/2};q)=q^{\pm\lambda/2}\tau_\lambda(q^{\mp 1}t;X_g;q)$, the solution is
\be\label{tauxq}
\tau_\lambda(t;X_{\e,\f};q)=\tau^{(\lambda)}_{x,1}(q)=A_1[\lambda]t\Big[(1+q^{-1}Ct\bar t)^{\lambda-1}\Big]_q
%\prod_{i=1}^{\lambda-1}\Big(1+q^{-1}Ct\bar t q^{\lambda-2i+1}\Big)
+A_2[\lambda]\bar t\Big[(1+qCt\bar t)^{\lambda-1}\Big]_q
%\prod_{i=1}^{\lambda-1}\Big(1+qCt\bar t q^{\lambda-2i+1}\Big)
\ee
and depends on two new arbitrary constants $A_1$ and $A_2$.
%Note that the shift of the upper limit in the product here corresponds to changing the degree $\lambda$ in (\ref{sol1}) to $\lambda-1$ in (\ref{sol2}).
The normalization $[\lambda]$ can not be absorbed into $A_1$ and $A_2$, because
in the two components of the product $\tau_\lambda\tau_{\lambda'}$
they are the same, while $[\lambda]$ and $[\lambda']$ are different.

Two more elements associated with two terms in the comultiplication sum are $X_g\cdot I/2$ and $X_g\cdot\h$. In this case, one has to insert $\tau_\lambda(t;X_g;q)\tau_{\lambda'}(t';X_x;q)+\tau_\lambda(t;X_x;q)\tau_{\lambda'}(t';X_g;q)$ into (\ref{sl2qeq}), and the solution to this equation is
\be
\tau_\lambda(t;X_x;q)=\tau^{(\lambda)}_{x,2}(q)=A_0\Big[(1+Ct\bar t)^\lambda\Big]_q
%\prod_{i=1}^{\lambda}\Big(1+Ct\bar t q^{\lambda-2i+1}\Big)
+A_3{\partial\over\partial C}\Big[(1+Ct\bar t)^\lambda\Big]_q
%\left[\prod_{i=1}^{\lambda}\Big(1+Ct\bar t q^{\lambda-2i+1}\Big)\right]
\ee
which also depends on two new arbitrary constants. Totally, one has four arbitrary constants, which suits the four constants $\alpha_x$, $A_1$, $A_2$, $A_3$ in equation (\ref{sol2}).

\bigskip

 One can proceed similarly
 for higher {\it ranks}, i.e. for higher elements $X_{xx}$  etc.

\bigskip

Let us note that, since the number of solution $\tau_\lambda(t;g;q)$ is rather restricted, one can not generate all $\tau_\lambda(t;X;q)$ just by taking its derivatives w.r.t. parameters of solution (\ref{sl2q}). However, missed $\tau_\lambda(t;X;q)$ can be generated by action on (\ref{tauq}) of the operators\footnote{For instance, one can generate (\ref{tauxq}) using that
\be
D_{t,q}\ \Big[(1+Ct\bar t)^\lambda\Big]_q=[\lambda]\bar t
\Big[(1+qCt\bar t)^{\lambda-1}\Big]_q\nn\\
D_{\bar t,q^{-1}}\ \Big[(1+Ct\bar t)^\lambda\Big]_q=[\lambda]t
\Big[(1+q^{-1}Ct\bar t)^{\lambda-1}\Big]_q\nn
\ee
} $D$, $\bar D$, hence, though not being related to the invariant B\"acklund transform, one can again say that {\bf $\tau_\lambda(t;X;q)$'s do not give rise, in a sense, to new hierarchies, similarly to the non-deformed case.}

Note also that equation (\ref{sl2qeq}) is looking much more involved as compared with the bilinear Hirota identities obtained using the intertwining operator approach \cite[Eqs.(29)-(30)]{GKLMM}. This is not that much surprising: the split Casimir operator is, roughly speaking, {\it quartic} in intertwining operators, while the operator $\Gamma$ in (\ref{Gamma})
was {\it quadratic} (see an explicit example in the next subsection).

\subsection{Solution associated with singular group elements}

Let us discuss the element $X_0$ describing the solution (\ref{tauX0}). We start with the limit $q \rightarrow 1$, and the corresponding solution is
\begin{equation}\label{ctauX0}
	\tau_\lambda(t;X_0)=\alpha_g\left(1+C_1 t\right)^\lambda (1+C_2 \bar t)^\lambda=
\alpha_g\left(1+C_1 t +C_2 \bar t + \underbrace{C_1 C_2}_{C_3} t \bar t \right)^\lambda
\end{equation}
This function though being a solution to the bilinear identities is not associated with a group element, since $C_3 \cdot 1 -  C_1 \cdot C_2$ is proportional to the determinant of the group element ($ ad-bc=1$!), which is non-zero, and, for (\ref{ctauX0}),  it is zero.

This is not that much surprising: we know that solutions singular from the representation theory point of view exist in integrable hierarchies. They typically are associated with a projector operator inserted. Such an operator depends on the representation. For instance, in the case of the standard KP/Toda hierarchy, when the $\tau$-function is associated with the fundamental representation, the singular solutions correspond to singular points of the infinite-dimensional Grassmannian, and the projector operator can be realized in terms of fermions \cite{KMMOZ}. 

In the case of the $\tau$-function (\ref{taucl}), the projector operator is $ P_{0}(\lambda):=\left| 0 \right \rangle_\lambda \! \left\langle 0 \right|$, i.e., in the representation $\lambda$, it is a projector on the highest weight vector $\left| 0 \right \rangle_\lambda$, and the element
\begin{equation}
X_0=e^{C_1 \f}e^{\alpha_1\h}  P_0 e^{\alpha_2\h} e^{C_2 \e} \,,
\end{equation}
 i.e. the projector splits the group element into two pieces associated with the two Borel subalgebras. Thus, the $\tau$-function becomes
\begin{equation}
\tau_\lambda(t;X_0)=\phantom{}_\lambda  {\left\langle 0 \right|} e^{t \e} e^{C_1 \f}e^{\alpha_1\h}   \left| 0 \right\rangle_\lambda \! \left\langle 0 \right| e^{\alpha_2\h} e^{C_2 \e} e^{\bar{t}  \f} \left| 0 \right \rangle_\lambda = e^{(\alpha_1+\alpha_2)\lambda} \left(1+C_1 t\right)^\lambda (1+C_2 \bar t)^\lambda
\end{equation}

Similarly, in the case of  generic $q$, i.e. of the $\tau$-function (\ref{tauq}),
\begin{equation}
	X_0=E_q(C_1 \f) e^{\alpha_1\h}  P_0 e^{\alpha_2\h} E_q(C_2 \e) \,,
\end{equation}
 where
\begin{equation}
	E_q(x) = \sum_{n=0}^{\infty} \dfrac{x^n}{[n]!} \, , \quad  \qquad  E_{q}(x)=E_{q^{-1}}(x)
\end{equation}
so that
\be
	%\boxed{
		\tau_\lambda(t;X_0;q)= \phantom{}_\lambda\left\langle 0 | e_{q}\left(t \e \right) E_q(C_1 \f) e^{\alpha_1\h}| 0 \right\rangle_\lambda \!  \left\langle 0 \right|e^{\alpha_2\h} E_q(C_2 \e) e_{q^{-1}}\left(\bar{t} \f  \right) \left| 0 \right \rangle_\lambda=\nn\\
=e^{(\alpha_1+\alpha_2)\lambda} \left\{\sum_{i=0}^\lambda q^{-i(i-1)/2}\binom{\lambda}{i}_q(C_1t)^i\right\}\times
\left\{\sum_{j=0}^\lambda q^{j(j-1)/2}\binom{\lambda}{j}_q(C_2\bar t)^j\right\}
	%}
\ee
reproducing (\ref{tauX0}).

This projector in matrix representations has the only non-zero matrix element: that at $i,j=1$. It can be also realized as a limit
\begin{equation}
	\lim_{\alpha \rightarrow \infty }q^{-\alpha \lambda}q^{\alpha\h} = \left| 0 \right \rangle_\lambda \! \left\langle 0 \right| = P_{0}(\lambda)
\end{equation}

The key reason why one can use the projector in constructing solutions to the bilinear equations is that {\bf the split Casimir commutes with the (tensor square of) the projector $P_0$}:
\begin{equation}
	\left[ P_0 \otimes P_0 , \Delta(\C_2) \right] =0
\end{equation}

\section{Infinite-dimensional hierarchies}

Now we consider two examples of infinite-dimensional hierarchies (which are basically known \cite{DJKM,Bourg}.

\subsection{Toda lattice hierarchy: $U(\mathfrak{sl}_N)$ algebra \label{slN}}

Let us consider the case of ${\cal G}=U(\mathfrak{sl}_N)$ at arbitrary $N$ keeping in mind the limit of $N\to\infty$, i.e. ${\cal G}=U(\mathfrak{sl}_\infty)$.
The $U_q(\mathfrak{sl}_N)$ algebra is given by the defining relations for the simple root generators $\e_i$, $f_i$ and the Cartan generators $\h_i$, $i=1,...,N-1$ (Chevalley basis),
\be
\phantom{.}[\e_i ,\f_j ]=\delta_{ij}\h_i\nn\\
\phantom{.}[\h_i ,\e_j ]=A_{ij}\e_j\\
\phantom{.}[\h_i ,\f_j ]=-A_{ij}\f_j\nn
\ee
where $A_{ij}$ is the Cartan matrix of the simple Lie algebra $\mathfrak{sl}_N$.
These relations has to be added by the Serre relations
\be
\hbox{ad}^{2}_{\e_i}(\e_j)=0,\ \ \ \ \ \ \ \ \ \ \hbox{ad}^{2}_{\f_i}(\f_j)=0,\ \ \ \ \ \ \ \ i\ne j
\ee
The second Casimir operator is
\be
\C_2=\sum_\Delta \Big(\e_\Delta\f_\Delta+\f_\Delta\e_\Delta\Big)+\sum_{i,j}^{N-1}A_{ij}^{-1}\h_i\h_j
\ee
where $\Delta$ denotes {\it all} roots, not only the simple ones so that $\f_\Delta$'s correspond to the positive root generators of the Lie algebra $\mathfrak{sl}_N$, and $\e_\Delta$'s, to the negative root generators.

The comultiplication is
\be
\Delta(\x)=\x\otimes I+I\otimes \x
\ee
where $\x$ is any element of $\e_\Delta$, $\f_\Delta$, $\h_i$. Thus, the second split Casimir operator is
\be\label{sCN}
\C=\Delta(\C_2)-I\otimes \C_2-\C_2\otimes I=\sum_\Delta \Big(\e_\Delta\otimes \f_\Delta+\f_\Delta\otimes\e_\Delta\Big)+\sum_{i,j}^{N-1}A_{ij}^{-1}\h_i\otimes\h_j
\ee
The standard $\tau$-function is defined \cite{DJKM,gentau}
\be
\tau_n(t;g)=\prescript{}{n}{\Big\langle} 0\Big|\exp\left(\sum_{k=1}^{N-1}t_k\E_k^{(n)}\right)g\exp\left(\sum_{k=1}^{N-1}\bar t_{k}\F_k^{(n)}\right)\Big|0\Big\rangle_n
\ee
where superscript $n$ refers to the $n$-th fundamental representation of $\mathfrak{sl}_N$, and $\E_k^{(n)}$, $\F_k^{(n)}$ are commutative operators that generate this representation, and $g$ is the group element. These operators are described in \cite{GKLMM,KMM,gentau}, in particular, $\E^{(1)}_k:=\Big(\sum_i\e_i\Big)^k$, $\F^{(1)}_k:=\Big(\sum_i\f_i\Big)^k$.
Considering fundamental representations is inevitable since the split Casimir operator (\ref{sCN}) contains $\e_\Delta$ with all $\Delta$, and, when pushing them to the right, one generates elements of the UEA that do not reduce to the degrees of $\F$, i.e. they can not be reproduced by differentiating. However, in the case of fundamental representations, these elements do reduce to the degrees of $\F$.

In order to understand it, we note that the fundamental representations admit a fermionic realization \cite{GKLMM,gentau}. Indeed, let us realize the simple root generators as ($i=1,\ldots,N-1$)
\be
\e_i=\psi_{i}\psi_{i+1}^\ast\nn\\
\f_i=\psi_{i+1}\psi_{i}^\ast\nn\\
\h_i=\psi_i\psi_{i}^\ast-\psi_{i+1}\psi^\ast_{i+1}
\ee
while all other generators $\e_\Delta$, $\f_\Delta$ are given by ($i,j=1,\ldots,N-1$)
\be
\e_{ij}=\psi_i\psi^\ast_j,\ \ \ \ \ i\langle j\nn\\
\f_{ij}=\psi_i\psi^\ast_j,\ \ \ \ \ i\rangle j
\ee
Here
\be
\phantom{.}\{\psi_i,\psi^\ast_j\}=\delta_{ij}\ \ \ \ \ \{\psi_i,\psi_j\}=0\ \ \ \ \ \{\psi^\ast_i,\psi^\ast_j\}=0
\ee
Then, one can rewrite the split Casimir operator (\ref{sCN}) in the form
\be\label{Cpsi}
\C=\sum_{i,j}\psi_i\psi^\ast_j\otimes\psi_j\psi^\ast_i-{1\over N+1}\Big(\sum_i\psi_i\psi^\ast_i\Big)\otimes
\Big(\sum_j\psi_j\psi^\ast_j\Big)=\nn\\
=\oint\oint{dz\over z}{dw\over w}\left[\psi(z)\psi^\ast(w)\otimes\psi(w)\psi^\ast(z)-
{1\over N+1}\psi(z)\psi^\ast(z)\otimes\psi(w)\psi^\ast(w)\right]
\ee
where
\be
\psi(z)=\sum_{i=1}^N\psi_iz^i,\ \ \ \ \ \ \ \psi^\ast(z)=\sum_{i=1}^N\psi^\ast_iz^{-i}
\ee
and the contour integrals go around origin\footnote{The integral measure includes the factor ${1\over 2\pi i}$.}.

Now let us introduce the vector $|0\rangle$ such that
\be
\psi_i^\ast|0\rangle=0\ \ \ \ \ i\ge 1
\ee
Then, the highest weight vector in the $n$-th fundamental representation is given by
\be
|0\rangle_n=\psi_n\psi_{n-1}\ldots\psi_2\psi_1|0\rangle:=|n\rangle
\ee
and one obtains that the first fundamental representation consists of the $N$ vectors \cite{GKLMM,KMM,gentau}
\be
|1\rangle=\psi_1|0\rangle\ \ \ \ \ \ \hbox{and}\ \ \ \ \ \ \F^{(1)}_{k-1}|1\rangle= \psi_k\psi_1^\ast|1\rangle=\psi_k|0\rangle,\ \ \ \ \ k=2,\ldots,N
\ee
or just
\be
\psi_k|0\rangle,\ \ \ \ \ k=1,\ldots,N
\ee
Similarly, the vector $\langle 0|$ at the same representation is defined as
\be
\langle 0|\psi_i=0\ \ \ \ \ i\ge 1,\ \ \ \ \ \ \ \langle n|=\langle 0|\psi_1\psi_2\ldots\psi_{n-1}\psi_n
\ee
so that
\be
\langle n|\E^{(1)}_{k-1}=\langle n| \psi_1\psi_k^\ast,\ \ \ \ \ k=2,\ldots,N
\ee
Similarly, the second fundamental representation consists of the $N(N-1)/2$ vectors
\be
\psi_k\psi_l|0\rangle\ \ \ \ \ \ N\ge k\rangle l\ge 1
\ee
etc.

In the fermionic terms, one can write the commutative operators generating the $n$-th fundamental representations in the universal form not depending on $n$:
\be
\E_k:=\sum_i\psi_i\psi_{i+k}^\ast,\ \ \ \ \ \ \F_k:=\sum_i\psi_{i+k}\psi_{i}^\ast
\ee
Note that \cite{DJKM,versus}
\be
\exp\left(\sum_kt_k\E_k\right)\psi_k\exp\left(-\sum_kt_k\E_k\right)=\sum\psi_{k-m}h_m(t)\nn\\
\exp\left(\sum_k\bar t_{k}\F_k\right)\psi_k\exp\left(-\sum_k\bar t_{k}\F_k\right)=\sum\psi_{k+m}h_m(\bar t)\nn\\
\exp\left(-\sum_kt_k\E_k\right)\psi_k^\ast\exp\left(\sum_kt_k\E_k\right)=\sum\psi_{k+m}^\ast h_m(t)\nn\\
\exp\left(-\sum_k\bar t_{k}\F_k\right)\psi_k^\ast\exp\left(\sum_k\bar t_{k}\F_k\right)=\sum\psi_{k-m}^\ast h_m(\bar t)
\ee
where $h_n(t)$ are the complete homogeneous symmetric polynomial of degree $n$ in variables $x_i$ taken as functions of the power sums\footnote{They can be obtained from the expansion $\exp\left(\sum_kt_{k}z^k\right)=\sum_kh_k(t)z^k$.} $t_k={1\over k}\sum_ix_i^k$. In other words,
\be
\exp\left(\sum_kt_k\E_k\right)\psi(z)\exp\left(-\sum_kt_k\E_k\right)=\left[\exp\left(\sum_kt_{k}z^k\right)\psi(z)\right]_{z,\leq N}\nn\\
\exp\left(\sum_k\bar t_{k}\F_k\right)\psi(z)\exp\left(-\sum_k\bar t_{k}\F_k\right)= \left[\exp\left(\sum_k\bar t_{k}z^{-k}\right)\psi(z)\right]_{z, > 0}\nn\\
\exp\left(-\sum_kt_k\E_k\right)\psi^\ast(z)\exp\left(\sum_kt_k\E_k\right)=\left[\exp\left(\sum_kt_{k}z^k\right)\psi^\ast(z)\right]_{z^{-1}, < 0}\nn\\
\exp\left(-\sum_k\bar t_{k}\F_k\right)\psi^\ast(z)\exp\left(\sum_k\bar t_{k}\F_k\right)=\left[\exp\left(\sum_k\bar t_{k}z^{-k}\right)\psi^\ast(z)\right]_{z^{-1},\le N}
\ee
where $[\ldots]_{z,\le N}$ means truncating the series in $z$ to have maximal degree $N$.

Finally, we need the formula that realizes the action of fermion on the highest vector state in terms of generators $\E_k$, $\F_k$ \cite{DJKM,Kharchev}:
\be
\psi(z)|n\rangle=z^{n+1}\exp\left(\sum_{k=1}^N{\F_k\over k}z^k\right)|n+1\rangle\ \ \ \ \ \ \ \psi^\ast(z)|n\rangle=z^{-n}\exp\left(-\sum_{k=1}^N{\F_k\over k}z^k\right)|n-1\rangle\nn\\
\langle n|\psi^\ast(z)=z^{-n-1}\langle n+1|\exp\left(\sum_{k=1}^N{\E_k\over k}z^{-k}\right)\ \ \ \ \ \
\langle n|\psi(z)=z^n\langle n-1|\exp\left(-\sum_{k=1}^N{\E_k\over k}z^{-k}\right)\
\ee
Using these formulas and formula (\ref{Cpsi}), one can derive the Hirota bilinear equations following the procedure described in sec.\ref{sl2}.

There is, however, a simpler way to get the Hirota bilinear equations: the intertwining operator approach of sec.\ref{inttw}. This is  the approach traditionally applied when dealing with the group elements, while, in the case of generic UEA elements, we used throughout the paper the approach via the split Casimir operators. In order to use the intertwining operator approach, we note that $\psi_k$, $\psi^\ast_k$ are the intertwining operators $E_R$ and $E_L$ and the element $\sum_k\psi_k\otimes\psi^\ast_k$ is just the element $\Gamma$, (\ref{Gamma}) which commutes with the comultiplication of the group element $g$, since \cite{DJKM}
\be\label{psir}
\psi_i g=\sum_j \alpha_{ij}g\psi_j,\ \ \ \ \ \ \ g\psi^\ast_i=\sum_j\alpha_{ji}\psi^\ast_j g
\ee
where $\alpha_{ij}$ are some numerical coefficients, and hence
\be\label{inCas}
\Big(\sum_i\psi_i\otimes\psi^\ast_i\Big)\Big(g\otimes g\Big)=
\Big(g\otimes g\Big)\Big(\sum_i\psi_i\otimes\psi^\ast_i\Big)
\ee
The properties (\ref{psir}) and, hence, (\ref{inCas}) are no longer correct when dealing with an arbitrary element $X$ of the UEA instead of the group one $g$:  {\bf  the bilinear combination $\sum_i \psi_i \otimes \psi^*_i$ is different from the split Casimir operator (\ref{Cpsi}), which is quartic in $\psi$}. However, what {\it is} correct is that the element $\sum_k\psi_k\otimes\psi^\ast_k$ commutes with the comultiplication $\Delta(\x)$, where $\x$ denotes any generator of the $\mathfrak{sl}_N$ Lie algebra (in the fundamental representation):
\be
\Big(\sum_i\psi_i\otimes\psi^\ast_i\Big)\Delta(\x)=\Big(\sum_i\psi_i\otimes\psi^\ast_i\Big)(\x\otimes I+I\otimes\x)=
\Big(\sum_i\psi_i\otimes\psi^\ast_i\Big)(\psi_k\psi^\ast_l\otimes I+I\otimes\psi_k\psi^\ast_l)=\nn\\
=\sum_i\psi_i\psi_k\psi^\ast_l\otimes\psi^\ast_i+\sum_i\psi_i\otimes\psi^\ast_i\psi_k\psi^\ast_l=
\sum_i\psi_k\psi^\ast_l\psi_i\otimes\psi^\ast_i+\sum_i\psi_i\otimes\psi_k\psi^\ast_l\psi^\ast_i=
\Delta(\x)\Big(\sum_i\psi_i\otimes\psi^\ast_i\Big)
\ee
for any $k,l=1,\ldots,N$.
This means that $\Gamma=\sum_k\psi_k\otimes\psi^\ast_k$ also commutes with $\Delta(X)$ where $X$ is an arbitrary element of UEA $U(\mathfrak{sl}_N)$, and the intertwining operator $\Gamma$ plays the same role in the product of two fundamental representations as the Casimir operator does (in arbitrary representations).

Thus, one can consider the element $\Gamma=\sum_k\psi_k\otimes\psi^\ast_k=\oint{dz\over z}\psi(z)\otimes\psi^\ast(z)$ instead of the split Casimir operator:
\be
\Big\langle n\Big|\exp\left(\sum_kt_k\E_k\right)\otimes \Big\langle m\Big|\exp\left(\sum_kt'_k\E_k\right)\Gamma\Delta(X)
\exp\left(\sum_k\bar t_k\F_k\right)\Big|n-1\Big\rangle\otimes\exp\left(\sum_k\bar t'_k\F_k\right)\Big|m+1\Big\rangle=\nn\\
=\Big\langle n\Big|\exp\left(\sum_kt_k\E_k\right)\otimes \Big\langle m\Big|\exp\left(\sum_kt'_k\E_k\right)\Delta(X)\Gamma
\exp\left(\sum_k\bar t_k\F_k\right)\Big|n-1\Big\rangle\otimes\exp\left(\sum_k\bar t'_k\F_k\right)\Big|m+1\Big\rangle
\ee
Hence, one obtains, instead of (\ref{besl2}),

{\footnotesize
\be\label{HeqT}
\sum_\alpha\oint{dz\over z}\left[\exp\left(-\sum_k\bar t_{k}z^{-k}\right)z^n\tau_n(t_k,\bar t_k+{z^k\over k};X_\alpha')\right]_{z,>0}
\left[\exp\left(\sum_k\bar t'_{k}z^{-k}\right)z^{-m-1}\tau_m(t'_k,\bar t'_k-{z^k\over k};X_\alpha'')\right]_{z^{-1},\le N}=\nn\\
=\sum_\alpha\oint{dz\over z}
\left[\exp\left(\sum_kt_{k}z^{k}\right)z^{n}\tau_{n-1}(t_k-{z^{-k}\over k},\bar t_k;X_\alpha')\right]_{z,\le N}
\left[\exp\left(-\sum_k t'_{k}z^k\right)z^{-m-1}\tau_{m+1}(t'_k+{z^{-k}\over k},\bar t'_k;X_\alpha'')\right]_{z^{-1},>0}
\nn
\ee
}

\noindent
Introducing the vertex operator
\be
V_n(t,z)=z^n\exp\left(\sum_kt_kz^k\right)\exp\left(-\sum_k{z^{-k}\over k}{\partial\over\partial t_k}\right)
\ee
one can rewrite
this relation in the form
\be
\sum_\alpha\oint {dz\over z}\left(\left[V_n(-\bar t,z^{-1})\tau_n(t,\bar t;X_\alpha')\right]_{z,>0}\cdot
\left[V_{-m-1}(\bar t',z^{-1})\tau_m(t',\bar t';X_\alpha'')\right]_{z^{-1},\le N}-\right.\nn\\
\left.-\left[V_n(t,z)\tau_{n-1}(t,\bar t;X_\alpha')\right]_{z,\le N}\cdot
\left[V_{-m-1}(-t',z)\tau_{m+1}(t',\bar t';X_\alpha'')\right]_{z^{-1},>0}\right)=0
\ee

This hierarchy at $N\to\infty$, i.e. for $U(\mathfrak{sl}_\infty)$, and when $X$ is a group element is nothing but the Toda lattice hierarchy \cite{DJKM,UT}. However, what we consider here is associated with the forced Toda lattice hierarchy \cite{Kaup}, which corresponds to the boundary condition $\tau_0=1$. The full Toda lattice hierarchy is described by the infinite number of fermions $\psi_k$, $\psi^\ast_k$, $k\in\mathbb{Z}$ instead of $k\in\mathbb{Z}_+$ as in this subsection, and the forced hierarchy can be embedded into the full one with a special choice of the group element (projection)  \cite{KMMOZ,versus}.

\subsection{Quantum toroidal algebra: $U_{q,t}(\Hat{\Hat{\mathfrak{gl}_1}})$}

As we observed in the previous section, in order to define a $\tau$-function associated with integrable flows, one typically needs to consider very peculiar representations, like the fundamental representations of the $\mathfrak{sl}_N$ algebra. The crucial property of these representations is that they are fully generated by a set of commuting operators ($\E_k$ and $\F_k$), and admit a fermionic, or bosonic realization.

Another example of this phenomenon is provided by the quantum toroidal $\mathfrak{gl}_1$ algebra \cite{Bourg}. However, in this case, in variance with the $\mathfrak{sl}_N$ algebra case, there are too few group elements in the UEA: most of them belong to the product of UEA and non-commutative algebra of functions on it. This is, however, a special story, to study {\it such} group elements in quantum toroidal algebra. Instead, it makes sense to lift the group element requitement and consider any elements of the UEA, at the price of having a large family of $\tau$-functions, as we explained above.

 This is what paper \cite{Bourg} just proposes to do. The peculiar representations in this case are the Fock representations, and counterparts of the fermions are known \cite{AFS,AKMMMMOZ}: they are associated with the operators intertwining three Fock representations (in terms of the quantum toroidal algebra, two of them are called horizontal and one, vertical). Accordingly, the intertwining operator $\Gamma$, which acts in the product of two (horizontal) representations is the screening charge of the $q$-Virasoro algebra \cite{AKMMMMOZ} acting in the product of two Fock representations and intertwining Fock representations with different weights.

Important is that the intertwining operators, the screening charges and the commutative operators giving the time flows $\E_k$ and $\F_k$ all admit bosonization: a representation in terms of the Heisenberg algebra. Then, similarly to sec.\ref{slN}, one can generate the Hirota bilinear identities in these terms \cite{Bourg}. This calculation requires a lot of technical detail about quantum toroidal algebra and its representations, they are well presented in \cite{Bourg}, thus, we do not reproduce it here, and simply refer the reader to that very nice paper for details.

There will be, however, an important comment, in the case of universal $\tau$-functions, which we will consider in the next section.

\section{On universal $\tau$-function}

The standard definition of the $\tau$-function is for the group element only. Hence, it is considered as an object in the algebra of functions, $\tau\in {\cal A}$. In the $q$-deformed case, the algebra of functions is non-commutative, and such are the $\tau$-functions. Let us briefly repeat the basic elements of the construction in this case \cite{KMM,gentau}.

Hence, we construct the group element given over
the non-commutative ring, the algebra of functions on the quantum group.
That is, we construct such an element
$g\in {\cal G}\otimes {\cal A}$ of the tensor product of
UEA ${\cal G}$ and its dual algebra of functions ${\cal A}$ that
\be\label{Tcoprod}
\Delta_U(g)=g\otimes_{U} g \in {\cal A}\otimes {\cal G}\otimes {\cal
G}
\ee
To construct this element \cite{FRT,FG,gentau,MV}, we fix some basis
$T^{(\alpha)}$ in ${\cal G}$. There exists a non-degenerated
pairing between ${\cal G}$ and
${\cal A}$, which we denote $\langle...\rangle$. We also fix the basis
$X^{(\beta)}$ in ${\cal A}$ orthogonal to $T^{(\alpha)}$ w.r.t. this
pairing. Then, the sum
\be\label{grouel}
\hbox{{\bf T}}\equiv\sum_{\alpha}X^{(\alpha)}\otimes T^{(\alpha)}\in {\cal A}
\otimes {\cal G}
\ee
is exactly the group element we are looking for. It is called the universal
{\bf T}-matrix (as it is intertwined by the universal
$R$-matrix) or the universal group element.

In order to prove that (\ref{grouel}) satisfies formula
(\ref{Tcoprod}) one should note that the matrices
$M^{\alpha\beta}_{\gamma}$ and $D^{\alpha}_{\beta\gamma}$ giving
respectively the multiplication and co-multiplication in
${\cal G}$
\be
T^{(\alpha)}\cdot T^{(\beta)}\equiv M^{\alpha\beta}_{\gamma}T^{(\gamma)},\ \
\Delta(T^{(\alpha)})\equiv D^{\alpha}_{\beta\gamma}T^{(\beta)}\otimes
T^{(\gamma)}
\ee
give rise to, inversely, co-multiplication and multiplication in the
dual algebra ${\cal A}$:
\be\label{DM}
D^{\alpha}_{\beta\gamma}=\left\langle \Delta(T^{(\alpha)}),X^{(\beta)}\otimes
X^{(\gamma)}\right\rangle\equiv \left\langle  T^{(\alpha)},X^{(\beta)}\cdot
X^{(\gamma)}\right\rangle\\
M^{\alpha\beta}_{\gamma}=\left\langle T^{(\alpha)}T^{(\beta)},X^{(\gamma)}\right\rangle=
\left\langle T^{(\alpha)}\otimes T^{(\beta)},\Delta(X^{(\gamma)})\right\rangle
\ee
Then,
\be
\Delta_U(\hbox{{\bf T}})=\sum_{\alpha}X^{(\alpha)}\otimes
\Delta_U(T^{(\alpha)})=
\sum_{\alpha,\beta,\gamma}D^{\alpha}_{\beta\gamma}X^{(\alpha)}\otimes
T^{(\beta)}\otimes T^{(\gamma)}
=\nn\\
=\sum_{\beta,\gamma}X^{(\beta)}X^{(\gamma)}
\otimes T^{(\beta)}\otimes T^{(\gamma)}=\hbox{{\bf T}}\otimes_U\hbox{{\bf T}}
\ee
This is the first defining property of the universal {\bf T}-operator,
which coincides with the classical one. The second property that allows
one to consider {\bf T} as an element of the ``true" group is the group
multiplication law $g\cdot g'=g''$ given by the map:
\be\label{grouplaw}
g\cdot g'\equiv \hbox{{\bf T}}\otimes_{\cal A}
\hbox{{\bf T}}\in {\cal A}\otimes {\cal A}\otimes {\cal G}
\longrightarrow g''\in {\cal A}\otimes {\cal G}
\ee
This map is canonically  given by the co-multiplication and is
again the universal {\bf T}-operator:
\be
\hbox{{\bf T}}\otimes_{\cal A}\hbox{{\bf T}}=\sum_{\alpha,\beta}X^{(\alpha)}\otimes
X^{(\beta)}\otimes T^{(\alpha)}T^{(\beta)}=\sum_{\alpha,\beta,\gamma}
M^{\gamma}_{\alpha,\beta}X^{(\alpha)}\otimes X^{(\beta)}\otimes T^{(\gamma)}=
\sum_{\alpha}\Delta(X^{(\alpha)})\otimes T^{(\alpha)}
\ee
i.e.
\be
g\equiv\hbox{{\bf T}}(X,T),\ \ g'\equiv \hbox{{\bf T}}(X',T),\ \
g''\equiv\hbox{{\bf T}}(X'',T)\nn\\
\ \ X\equiv\{X^{(\alpha)}\otimes I\}\in {\cal A}\otimes I,
\ \ X'\equiv\{I\otimes X^{(\alpha)}\}\in I\otimes {\cal A}\nn\\
X''\equiv\{\Delta(X^{(\alpha)})\}\in {\cal A}\otimes {\cal A}
\ee

One of the possibilities to deal with such $\tau$-functions, which are no longer commutative is to consider concrete representations of ${\cal A}$. In \cite{Bourg}, the authors suggest to consider instead of an element $\xi\in{\cal A}$, the corresponding dual element from UEA ${\cal G}$ so that, in the definition of the $\tau$-function, one replaces the universal group element $T$ with the universal $R$-matrix. The object obtained this way is still non-commutative. Its advantage is that the representations of UEA and the corresponding $R$-matrices are better studied, and the drawback is that, in this case, the $\tau$-function has no a natural $q\to 1$ limit that would reproduce the standard hierarchy.

\section{Conclusion}

The goal of this paper was to analyze possible implications of the change of the basic idea
behind the bilinear Hirota equations, which describe comultiplication in Hopf algebras
in terms of the generating ($\tau$-)functions of matrix elements.
The problem is that the standard formalism of \cite{GKLMM} requires $\tau$-functions to
be made from the {\it group elements}, satisfying $\Delta(g) = g\otimes g$,
which forces them to take values in the
algebra of functions, the latter being non-commutative in the case of quantum groups.
The lack of canonical coordinatization of such algebras and underdeveloped theory of non-commutative
special functions make Hirota equations in these cases badly looking and distractive.
This is a big problem because of increasing role of quantum groups and, especially, quantum toroidal algebras
in the modern generalizations of matrix models for the purposes of string/brane physics.

We investigate an alternative approach, avoiding use of the {\it group elements},
and working directly with the $c$-number matrix elements, which attracted new attention
after a recent paper by J.-E. Bourgine and A. Garbali \cite{Bourg}.
The price to pay here is that the bilinear equations do not close and one needs to introduce
many different $\tau$-functions associated with different flows along the ``Universal Grassmannian"
(actually, the elements of the universal enveloping algebra).
The set of Hirota equations also increases and describes interrelations between these
$\tau$-functions.
Instead they deal with the ordinary $c$-number functions and, in this sense, are more familiar
and more comprehensible.
We consider a number of simple examples.

At this moment, it is still difficult to choose between the two approaches, the fundamental
but technically difficult one with the non-commutative {\it group elements},
and technically transparent one with $c$-number $\tau$-functions,
which instead does not fully reflect ``physics of the problem" and provides {\it systems} of equations
with complicated solutions.
It looks like both require some attention.
Hopefully the very existence of the two competitive approaches would attract more attention to this problem,
and we will finally obtain a healthy and broadly recognized extension of integrability theory
to the field of quantum groups and DIM algebras.

\section*{Acknowledgements}

This work was partly supported by grants RFBR 21-51-46010-ST-a and 21-52-52004-MNT-a, and by the grants of the
Foundation for the Advancement of Theoretical Physics and Mathematics ``BASIS".

\end{document}